\documentclass[a4paper,11pt]{article}
\pdfoutput=1 

\usepackage{jheppub_modified} 

\usepackage[T1]{fontenc} 

\usepackage{verbatim}
\usepackage{color}
\usepackage{graphicx}
\usepackage{subcaption}

\usepackage{bigcenter}

\usepackage{sectsty}
\allsectionsfont{\boldmath}

\usepackage{xspace}
\usepackage[dvipsnames,table]{xcolor}
\usepackage{tabularx}
\newcolumntype{C}[1]{>{\centering\arraybackslash}p{#1}}

\newcommand{\eq}[1]{Eq.~(\ref{#1})}
\newcommand{\bib}[1]{Ref.~\cite{#1}}

\newcommand{\bibs}[1]{\cite{#1}}
\newcommand{\fig}[1]{Fig.~\ref{#1}}
\newcommand{\tab}[1]{Table~\ref{#1}}

\newcommand{\sect}[1]{Section~\ref{#1}}

\newcommand{\appen}[1]{Appendix~\ref{#1}}

\newcommand{\bea}{\begin{eqnarray}}
\newcommand{\eea}{\end{eqnarray}}

\newcommand{\crn}{\nonumber \\}

\newcommand{\fr}{\frac}

\newcommand{\gev}{{\unskip\,\text{GeV}}}

\title{NLO electroweak corrections to doubly-polarized $W^+W^-$ production at the LHC}

\author[a]{Thi Nhung Dao,}
\author[a]{Duc Ninh Le}

\affiliation[a]{Faculty of Fundamental Sciences, PHENIKAA University, Hanoi 12116, Vietnam}

\emailAdd{nhung.daothi@phenikaa-uni.edu.vn}
\emailAdd{ninh.leduc@phenikaa-uni.edu.vn}

\abstract{We present new results of next-to-leading order (NLO) electroweak corrections to 
doubly-polarized cross sections of $W^+W^-$ production at the LHC. The calculation is performed 
for the leptonic final state of $e^+\mu_e \mu^- \bar{\nu}_\mu$ using the double-pole approximation in the 
diboson center-of-mass frame. 
NLO QCD corrections and subleading contributions from the $gg$, $b\bar{b}$, $\gamma\gamma$ induced processes 
are taken into account in the numerical results. We found that NLO EW corrections are small for angular distributions but 
can reach tens of percent for transverse momentum distributions at high energies, e.g. reaching $-40\%$ at $p_{T,e}\approx 300$ GeV. 
In these high $p_T$ regions, EW corrections are largest for the doubly-transverse mode.}

\begin{document}
\maketitle
\flushbottom

\section{Introduction}
\label{sect:intro}
Measuring polarized cross sections of diboson production processes has been being performed from the LEP \cite{OPAL:1998ixj} to the LHC. 
Recent results include $W^\pm Z$ measurements at ATLAS \cite{Aaboud:2019gxl,ATLAS:2022oge} and at CMS \cite{CMS:2021icx},  
same-sign $WWjj$ at CMS \cite{CMS:2020etf}, and $ZZ$ at ATLAS \cite{ATLAS:2023zrv}. 

From the theoretical side, next-to-leading order (NLO) QCD predictions for doubly-polarized cross sections 
have been provided for $W^+W^-$ \cite{Denner:2020bcz}, $W^\pm Z$ \cite{Denner:2020eck,Le:2022lrp,Le:2022ppa,Dao:2023pkl}, and $ZZ$ \cite{Denner:2021csi}.
NLO electroweak (EW) corrections are available for $W^\pm Z$ \cite{Le:2022lrp,Le:2022ppa,Dao:2023pkl} and $ZZ$ \cite{Denner:2021csi}, 
for fully leptonic decays. 
Next-to-next-to-leading order (NNLO) QCD results have been obtained for $W^+W^-$ \cite{Poncelet:2021jmj}. 
Very recently, the calculation of polarized cross sections for multi-boson production processes 
has been made possible with SHERPA \cite{Hoppe:2023uux}, allowing for simulations up to the level of approximate fixed-order 
NLO QCD corrections matched with parton shower. For the inclusive diboson processes, 
the full NLO QCD amplitudes have been combined with parton-shower effects in the POWHEG-BOX framework 
in \cite{Pelliccioli:2023zpd}. Going beyond the fully leptonic decays, NLO QCD results for the $WZ$ production with 
semileptonic decays have been obtained in \cite{Denner:2022riz}.

In this paper, continuing our previous works for the $W^\pm Z$ processes, 
we present new results of NLO EW corrections to the $W^+W^-$ production in fully leptonic final states at the LHC. 
Subleading contributions from the $b\bar{b}$, $gg$, and $\gamma\gamma$ processes are calculated at leading order (LO) and 
will be included in the final results.  
The numerical predictions provided here are not state-of-the-art, 
as NNLO QCD corrections and parton-shower effects are not incorporated.

The paper is organized as follows. In \sect{sect:pol_LO} we explain how the doubly-polarized cross sections are calculated at 
LO and provide here the definition of various polarization modes. In \sect{sect:pol_NLO}, the method to calculate NLO QCD and EW corrections is briefly explained, leaving the new technical details for \appen{appen_NLO_cal}. 
Information on the $b\bar{b}$, $gg$, and $\gamma\gamma$ processes is given in \sect{sect:pol_gba}. 
Numerical results for integrated cross sections and kinematic distributions are presented in \sect{sect:results}. 
We conclude in \sect{sect:conclusion}. 
\section{Doubly-polarized cross sections at LO}
\label{sect:pol_LO}
\begin{figure}[ht!]
  \centering
  \includegraphics[width=0.8\textwidth]{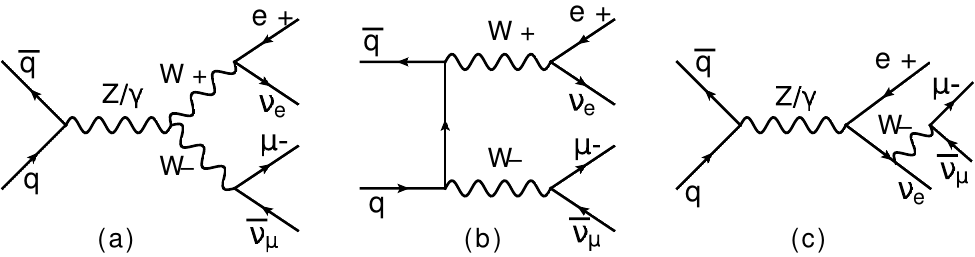}
  \caption{Representative Feynman diagrams at leading order. Diagrams (a) and 
  (b) are included in the DPA, while (c) is excluded.}
  \label{fig:LO_diags}
\end{figure}

In this work, we consider the process
\bea
p(k_1) + p(k_2) \to e^{+}(k_3) + \nu_e(k_4) + \mu^{-}
(k_5) + \bar{\nu}_\mu(k_6) + X,
\label{eq:proc1}
\eea
whose Feynman diagrams at leading order are shown in \fig{fig:LO_diags}.

To measure the doubly-resonant signal, we focus on 
the kinematical region where the final-state leptons mainly 
come from the on-shell (OS) $W$ bosons. In order to separate this contribution 
in a gauge invariant way, we use the double-pole approximation (DPA) \cite{Aeppli:1993cb,Aeppli:1993rs,Denner:2000bj}, where only the 
doubly-resonant diagrams (i.e. diagrams (a) and (b) in \fig{fig:LO_diags}) are selected. 
Non-doubly-resonant diagrams, such as diagram (c) in \fig{fig:LO_diags}, are excluded. 
This method has been used 
in recent theoretical works $ZZ$ \cite{Denner:2021csi}, $WW$ \cite{Denner:2020bcz}, $WZ$ \cite{Denner:2020eck,Le:2022ppa}. 
These results were then used by 
ATLAS in their new $WZ$ \cite{ATLAS:2022oge} and $ZZ$ \cite{ATLAS:2023zrv} measurements.  

At leading order, the doubly-resonant contribution to the process \eq{eq:proc1} is 
calculated from
\bea
p(k_1) + p(k_2) \to W^+(q_1) + W^-(q_2) \to e^{+}(k_3) + \nu_e(k_4) + \mu^{-}
(k_5) + \bar{\nu}_\mu(k_6) + X.
\label{eq:proc1_VV}
\eea
In the DPA, the process can be viewed as an OS production of $W^+W^-$ followed by the OS
decays $W^+\to e^+ \nu_e$ and $W^-\to\mu^-\bar{\nu}_\mu$. Spin correlations 
between the production and decays are fully taken into account. 

To be more specific, the DPA amplitude at LO is calculated as, writing $V_1 = W^+$, $V_2 = W^-$, 
$l_1 = e^+$, $l_2 = \nu_e$, $l_3 = \mu^-$, $l_4 = \bar{\nu}_\mu$,
\bea
\mathcal{A}_\text{LO,DPA}^{\bar{q}q\to V_1V_2\to 4l} = \fr{1}{Q_1Q_2}
\sum_{\lambda_1,\lambda_2=1}^{3}
\mathcal{A}_\text{LO}^{\bar{q}q\to V_1V_2}(\hat{k}_i,\lambda_1,\lambda_2)\mathcal{A}_\text{LO}^{V_1\to
    l_1l_2}(\hat{k}_i,\lambda_1)\mathcal{A}_\text{LO}^{V_2\to l_3l_4}(\hat{k}_i,\lambda_2)
,\label{eq:LO_DPA}
\eea
with 
\bea
Q_j = q_j^2 - M_{V_j}^2 + iM_{V_j}\Gamma_{V_j}\;\; (j=1,2),
\label{eq:Qi_def}
\eea
where $q_1 = k_3+k_4$, $q_2 = k_5 + k_6$, $M_{V_j}$ and $\Gamma_{V_j}$ are the
physical mass and width of the gauge boson $V_j$, and $\lambda_j$ are the
polarization indices of the gauge bosons. The helicity indices of the initial-state quarks and final-state leptons are implicit, 
meaning that the full helicity amplitude on the l.h.s. $\mathcal{A}_\text{LO,DPA}^{\bar{q}q\to V_1V_2\to 4l}$ 
stands for 
$\mathcal{A}_\text{LO,DPA}^{\bar{q}q\to V_1V_2\to 4l}(\sigma_{\bar{q}},\sigma_q,\sigma_{l_1},\sigma_{l_2},\sigma_{l_3},\sigma_{l_4})$ 
with $\sigma_{\bar{q}}$, $\sigma_q$ 
being the helicity indices of the initial-state quarks; $\sigma_{l_1}$, $\sigma_{l_2}$, $\sigma_{l_3}$, $\sigma_{l_4}$
of the final-state leptons. 
Correspondingly, on the r.h.s. we have $\mathcal{A}_\text{LO}^{\bar{q}q\to V_1V_2} = \mathcal{A}_\text{LO}^{\bar{q}q\to V_1V_2}(\sigma_{\bar{q}},\sigma_q)$, $\mathcal{A}_\text{LO}^{V_1\to
    l_1l_2}=\mathcal{A}_\text{LO}^{V_1\to
    l_1l_2}(\sigma_{l_1},\sigma_{l_2})$, 
    $\mathcal{A}_\text{LO}^{V_2\to l_3l_4}=\mathcal{A}_\text{LO}^{V_2\to l_3l_4}(\sigma_{l_3},\sigma_{l_4})$. 
    The squared amplitude then automatically includes correlations between different helicity states of the final leptons. 

A key point is that all helicity amplitudes $\mathcal{A}$ on
the r.h.s. are calculated using OS momenta $\hat{k}_i$ for the
final-state leptons as well as OS momenta $\hat{q}_j$ for the
intermediate $W$ bosons, to obtain a gauge-invariant result. 
The OS momenta $\hat{k}_i$ are computed from the off-shell momenta $k_i$ via an OS mapping.  
This mapping is not unique, however the differences between different choices are very small, 
of order $\alpha \Gamma_V/(\pi M_V)$ \cite{Denner:2000bj}, hence of no practical importance. 
In this work, we use the same OS mappings as in \bib{Le:2022ppa}. 
A necessary condition for the existence of OS mappings is that the 
invariant mass of the four-lepton system must be greater than $2M_W$, 
which is required at LO and NLO. 

From \eq{eq:LO_DPA} we can define the doubly-polarized cross sections as follows. 
A $W$ boson has three physical polarization states \footnote{In a general gauge choice, a fourth 
state is introduced. Its contribution is proportional to the masses of the decay products, which are 
neglected in this work.}: two transverse modes $\lambda = 1$ and
$\lambda = 3$ and one longitudinal $\lambda = 2$. 
The intermediate $WW$ system has therefore $9$ polarization states in total. The unpolarized 
amplitude defined in \eq{eq:LO_DPA} is the sum of these $9$ polarized amplitudes. 
The unpolarized cross section is then divided into the following five contributions:
\begin{itemize}
\item $W^+_L W^-_L$: longitudinal-longitudinal (LL) contribution, obtained with selecting $\lambda_1=\lambda_2=2$
  in the sum of \eq{eq:LO_DPA}. In addition, the sum over the helicities of the the initial-state quarks and 
  final-state leptons is performed.  
\item $W^+_L W^-_T$: longitudinal-transverse (LT), obtained with $\lambda_1=2$, 
$\lambda_2=1,3$. Note that the LT cross section includes the interference (correlation) between the $(21)$ and $(23)$ amplitudes.
\item $W^+_T W^-_L$: transverse-longitudinal (TL), obtained with $\lambda_1=1,3$, 
$\lambda_2=2$. The interference between the $(12)$ and $(32)$ amplitudes is here included.
\item $W^+_T W^-_T$: transverse-transverse (TT), obtained with $\lambda_1=1,3$, 
$\lambda_2=1,3$. The interference terms between the $(11)$, $(13)$, $(31)$, $(33)$ amplitudes are here included.
\item Interference: The difference between the unpolarized cross section and the sum of the above four contributions. 
This includes the interference between the above LL, LT, TL, TT amplitudes.     
\end{itemize}
For the NLO QCD and EW corrections, the definition of double-pole amplitudes 
need include the virtual corrections, the gluon/photon induced and radiation processes. 
This issue is addressed next.
\section{Doubly-polarized cross sections at NLO QCD+EW}
\label{sect:pol_NLO}
We use the same method as described in \cite{Le:2022ppa} for the $WZ$ process, hence will refrain from repeating technical details here. 
This section is therefore to highlight the new ingredients occurring in the $W^+W^-$ process. 
It is important to note that NLO QCD+EW corrections are calculated only for the $q\bar{q}\to 4l$ processes with $q=u,d,c,s$. 
The $\gamma\gamma$ and $b\bar{b}$ contributions are much smaller and hence computed only at LO.

The NLO QCD calculation was first provided in \cite{Denner:2020bcz}. This calculation is the same 
for all diboson processes. Although the main goal of this work is to calculate the NLO EW corrections, 
we have re-calculated the NLO QCD contribution so that our numerical results can be used for phenomenological studies. 

For the NLO EW corrections, compared to the $WZ$ process, there are two new ingredients: the final-state emitter and 
final-state spectator contribution of the OS production part in the dipole-subtraction method, and the $\gamma\gamma \to 4l$ 
process as a second underlying-Born contribution (besides the $q\bar{q}$ underlying-Born channel which is 
the only contribution in the $WZ$ case) to the quark-photon induced real-emission processes occurring at NLO. 
It suffices here to say that these calculations are fairly straightforward by following the guidelines provided in \cite{Le:2022ppa}.    
Since this is rather technical and deeply related to the dipole-subtraction method \cite{Catani:1996vz,Dittmaier:1999mb}, 
we refer the reader to \appen{appen_NLO_cal} for calculational details. 

We note that the NLO DPA calculation presented here is based on the dipole-subtraction formalism, where the subtraction terms 
for the case of massive external-state particles are available (needed for the OS production and decay parts). 
It would be interesting to do the calculation using a different 
subtraction method such as the Frixione-Kunszt-Signer sector-subtraction scheme \cite{Frixione:1995ms}.  
\section{$gg$, $b\bar{b}$ and $\gamma\gamma$}
\label{sect:pol_gba}
\begin{figure}[ht!]
  \centering
  \includegraphics[width=0.8\textwidth]{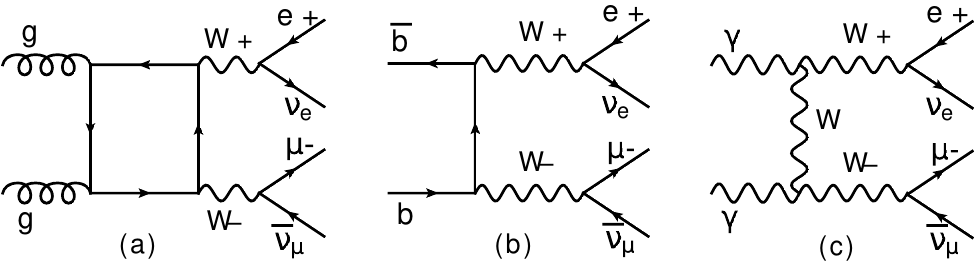}
  \caption{Representative DPA Feynman diagrams at leading order for the loop-induced $gg$ fusion (a), $b\bar{b}$ annihilation (b) 
  and $\gamma\gamma$ (c).}
  \label{fig:LO_diags_gba}
\end{figure}
For the sake of completeness, subleading contributions from the loop-induced $gg$, 
$b\bar{b}$ and $\gamma\gamma$ processes are also taken into account. 
Their 
representative Feynman diagrams are depicted in \fig{fig:LO_diags_gba}. 
They are calculated at LO using the DPA 
as described in \sect{sect:pol_LO}. 
Since $m_{4l} > 2M_W$, a DPA requirement so that both gauge bosons can be simultaneously on-shell, 
the Higgs resonance in the $gg$ process is suppressed. 
The bottom mass is neglected in the $b\bar{b}$ case, while we use a finite value $m_b=4.7\gev$ for the $gg$ 
contribution. Neglecting the bottom mass changes the $gg$ cross sections by $+0.3\%$, $-0.7\%$, $+0.4\%$, 
$+0.5\%$, $-0.05\%$ for the unpolarized, LL, LT, TL, TT cases, respectively, using the input parameters given 
in \sect{sect:results}. 

The NLO corrections for the $\gamma\gamma$ process can be straightforwardly calculated using the same method for 
the $q\bar{q}$ processes. We however refrain from including them here to reduce the computing time, because they are very small. 
Even the largest corrections, occurring in the region of large $p_{T,W}$ or of large $m_{WW}$, are found 
to be small for the unpolarized on-shell production \cite{Baglio:2013toa} (see Fig. 15 there). 
Moreover, there is no reason to expect EW corrections to be large for any particular polarization mode in this channel.      

It is expected that the 
unpolarized cross section is largest for the $gg$ channel, 
while the $b\bar{b}$ and $\gamma\gamma$ ones are much smaller due to small values of the 
parton distribution functions (PDF). 
However, it is not obvious whether this hierarchy of the unpolarized 
cross sections holds true for the polarized ones. 
It turns out that the $b\bar{b}$ contribution to the LL cross section is 
much larger than the $gg$ one. We will discuss this more in the numerical result section. 
This is an indication that NLO QCD and EW corrections to the $b\bar{b}$ may be important when high accuracy is required. 
In this context, we remind the reader of the $bg$ and $b\gamma$ real-emission processes occurring at NLO, 
which bring in a new mechanism related to an intermediate on-shell (anti)top quark. 
While these are genuine NLO corrections to the $W^+W^-$ signal process, they belong also 
to the single-top irreducible background. Similarly, $t\bar{t}$ production contributes to both the signal 
(starting from the NNLO level) and the irreducible background.  
The best strategy is therefore to minimize their size 
by using $b$ tagging. 
Since this is a rather delicate issue which highly depends on the experimental analysis 
and is not the purpose of this work, 
we will leave it for future work needing a careful investigation.   
\section{Numerical results}
\label{sect:results}
We use the same set of input parameters and renormalization schemes for NLO QCD and NLO EW 
calculations as in \bib{Le:2022ppa} for this work. 
The only difference is that $m_b=4.7\gev$ for the loop-induced gluon gluon contribution, 
otherwise the bottom quark is massless. 
Results will be presented for the LHC at $13$ TeV center-of-mass energy. 
The factorization and renormalization scales are chosen at a fixed value 
$\mu_F = \mu_R = M_W$, where $M_W=80.385\,\gev$. 
The parton distribution functions are obtained using the Hessian set 
{\tt
  LUXqed17\char`_plus\char`_PDF4LHC15\char`_nnlo\char`_30}~\bibs{Watt:2012tq,Gao:2013bia,Harland-Lang:2014zoa,Ball:2014uwa,Butterworth:2015oua,Dulat:2015mca,deFlorian:2015ujt,Carrazza:2015aoa,Manohar:2016nzj,Manohar:2017eqh} via the library {\tt LHAPDF6}~\bibs{Buckley:2014ana}. For the NLO EW corrections, we use the DIS scheme \cite{Dittmaier:2009cr} for the PDF counterterms. 
  The difference compared to the $\overline{MS}$ scheme \cite{Dittmaier:2009cr} is negligible.   

For NLO EW corrections, an additional photon 
can be emitted. Before applying real analysis cuts on the charged leptons, we 
do lepton-photon recombination to define a dressed lepton. A dressed lepton 
is defined as $p'_\ell = p_\ell + p_\gamma$ if $\Delta
R(\ell,\gamma) \equiv \sqrt{(\Delta\eta)^2+(\Delta\phi)^2}< 0.1$, i.e. when the photon 
is close enough to the bare lepton.   
In case two charged leptons both satisfy this condition, 
the closest one to the photon is chosen. Only photons with $|y_\gamma|<5$ are eligible for this 
recombination, as they are considered lost in the beam pipe otherwise. 
The letter $\ell$ can be either $e$ or $\mu$ 
and $p$ denotes momentum in the laboratory (LAB) frame.
Finally, the ATLAS fiducial phase-space cuts are applied as follows 
\begin{align}
        & p_{T,\ell} > 27\gev, \quad p_{T,\text{miss}} > 20\gev, \quad |\eta_\ell|<2.5, \quad m_{e\mu} > 55\gev,\crn
        & \text{jet veto (no jets with $p_{T,j}>35\gev$ and $|\eta_j|<4.5$)},
\end{align}
where the jet veto is used to suppress the top-quark and $Wj$ (which contributes when a nonprompt charged lepton is 
radiated off the jet) backgrounds. 
This set of cuts is used in \cite{Denner:2020bcz}, which is adapted from \cite{ATLAS:2019rob}. 
It is worth noting here that the CMS $W^+W^-$ analyses in \bib{CMS:2020mxy} do not use a jet veto. 

Apart from these cuts, there is an additional requirement of $m_{4l} > 2M_W$, as above mentioned, 
so that events with two on-shell $W$ bosons are available. If one needs to estimate off-shell effects for 
unpolarized cross sections, many tools are available in the market, see e.g. VBFNLO~\cite{Baglio:2014uba} for full NLO QCD 
calculation, \bib{Biedermann:2016guo} for full NLO EW, \bib{Grazzini:2019jkl} for full NNLO QCD + NLO EW.

Before presenting our numerical results, we define here various terminologies used in the next sections. 
LO results include only the $q\bar{q}$ contributions with $q=u,d,c,s$. 
NLO QCD, NLO EW and NLO QCD+EW (also written as QCDEW for short) results include additionally virtual and real-emission 
processes (e.g. $gq$ and $\gamma q$ initial states) related only to those processes.  
The subscript {\em all} then means the sum of the NLO QCDEW and $gg$, $b\bar{b}$, $\gamma\gamma$ contributions. 

To quantify various effects, we define $\bar{\delta}_\text{EW}$ as the ratio of the NLO EW correction (i.e. without the LO contribution) 
over the NLO QCD cross section. Similarly, $\bar{\delta}_{gg}$, $\bar{\delta}_{b\bar{b}}$ and $\bar{\delta}_{\gamma\gamma}$ 
are defined with respect to the same denominator to facilitate comparisons between those effects. 

Numerical results for polarized cross sections will be presented for the VV center-of-mass frame, called VV frame for short. 
This frame was used in CMS same-sign $WWjj$ measurement \cite{CMS:2020etf} and ATLAS $WZ$ measurement \cite{ATLAS:2022oge}. 
Theoretical study in \cite{Denner:2020eck} for $WZ$ production shows that polarization fractions differ significantly between 
the VV and laboratory frames, in particular the LL fraction is larger in the VV frame. The parton-parton
center-of-mass frame was also used in \cite{CMS:2020etf}.

We have performed several checks on our results. At the unpolarized level, comparison to the NLO QCD results of 
\cite{Denner:2020bcz} has been done, the difference is less than $0.1\%$. For the polarized cross sections calculated in 
the VV frame, the same polarization separation procedure as done for the $WZ$ process is used here. 
This procedure has been well-checked as our $WZ$ polarized cross sections agree with the ones of \cite{Denner:2020eck} at the NLO QCD level, the differences are less than $0.1\%$. 
For the NLO EW cross sections, various consistency checks including UV and IR finiteness have been performed.     
\subsection{Integrated polarized cross sections}
\label{sect:XS}
We first show results for the polarized (LL, LT, TL, TT) and unpolarized integrated cross sections in \tab{tab:xs_fr}. 
The interference, calculated by subtracting the polarized cross sections from the unpolarized one, is shown in the bottom row. 
To facilitate comparisons with our results, the cross sections are provided at LO, NLO QCD, NLO QCDEW.  
The best values, the sum of the NLO QCDEW, $gg$, $b\bar{b}$ and $\gamma\gamma$ contributions, 
are shown in the column $\sigma_\text{all}$. 
Besides, various corrections $\bar{\delta}_\text{EW}$, $\bar{\delta}_{gg}$, $\bar{\delta}_{b\bar{b}}$ and $\bar{\delta}_{\gamma\gamma}$
are also presented. The last column is the polarization fraction calculated as 
$f^i_\text{all} = \sigma^i_\text{all}/\sigma^\text{unpol.}_\text{all}$ with $i=$ LL, LT, TL, TT, interference. 
The scale uncertainties are computed by varying $\mu_F$ and $\mu_R$ independently in the range from 
$M_W/2$ to $2M_W$ following the well-known seven-point method (see e.g. \cite{Le:2022lrp} for more details).   
\begin{table}[h!]
 \renewcommand{\arraystretch}{1.3}
\begin{bigcenter}
\setlength\tabcolsep{0.03cm}
\fontsize{8.0}{8.0}
\begin{tabular}{|c|c|c|c|c|c|c|c|c|c|}\hline
  & $\sigma_\text{LO}\,\text{[fb]}$ & $\sigma^\text{QCD}_\text{NLO}\,\text{[fb]}$ & $\sigma^\text{QCDEW}_\text{NLO}\,\text{[fb]}$ & $\sigma_\text{all}\,\text{[fb]}$ & $\overline{\delta}_\text{EW}\,\text{[\%]}$ & $\overline{\delta}_{gg}\,\text{[\%]}$ & $\overline{\delta}_{b\overline{b}}\,\text{[\%]}$ & $\overline{\delta}_{\gamma\gamma}\,\text{[\%]}$ & $f_\text{all}\,\text{[\%]}$ \\
\hline
{\fontsize{7.0}{7.0}$\text{Unpolarized}$} & $198.14(1)_{-6.5\%}^{+5.3\%}$ & $210.91(3)_{-2.2\%}^{+1.6\%}$ & $202.90(3)_{-1.9\%}^{+1.3\%}$ & $222.41(3)_{-2.5\%}^{+2.2\%}$ & $-3.80$ & $6.20$ & $1.87$ & $1.18$ & $100$\\
\hline
{\fontsize{7.0}{7.0}$W^{+}_{L}W^{-}_{L}$} & $12.99_{-7.4\%}^{+6.1\%}$ & $14.03_{-2.6\%}^{+1.9\%}$ & $13.64_{-2.4\%}^{+1.7\%}$ & $16.46_{-5.7\%}^{+4.7\%}$ & $-2.75$ & $4.08$ & $15.11$ & $0.94$ & $7.4$\\
{\fontsize{7.0}{7.0}$W^{+}_{L}W^{-}_{T}$} & $21.67_{-7.5\%}^{+6.3\%}$ & $24.86_{-2.6\%}^{+1.8\%}$ & $24.28_{-2.5\%}^{+1.7\%}$ & $25.75_{-3.5\%}^{+2.6\%}$ & $-2.32$ & $1.56$ & $3.86$ & $0.50$ & $11.6$\\
{\fontsize{7.0}{7.0}$W^{+}_{T}W^{-}_{L}$} & $22.14_{-7.5\%}^{+6.2\%}$ & $25.56_{-2.6\%}^{+1.8\%}$ & $24.96_{-2.5\%}^{+1.7\%}$ & $26.43_{-3.5\%}^{+2.6\%}$ & $-2.34$ & $1.52$ & $3.75$ & $0.48$ & $11.9$\\
{\fontsize{7.0}{7.0}$W^{+}_{T}W^{-}_{T}$} & $140.44_{-6.0\%}^{+4.8\%}$ & $144.97(2)_{-1.9\%}^{+1.6\%}$ & $138.42(2)_{-1.6\%}^{+1.4\%}$ & $152.95(3)_{-1.9\%}^{+2.3\%}$ & $-4.52$ & $8.32$ & $0.25$ & $1.46$ & $68.8$\\
\hline
{\fontsize{7.0}{7.0}$\text{Interference}$} & $0.90(1)$ & $1.50(4)$ & $1.60(4)$ & $0.81(4)$ & $--$ & $--$ & $--$ & $--$ & $0.4$\\
\hline
\end{tabular}
\caption{\small Unpolarized and doubly polarized cross sections in fb
  calculated in the VV frame for the process $p p \to W^+ W^-\to e^+ \nu_e \mu^- \bar{\nu}_\mu + X$.   
  The statistical uncertainties (in parenthesis) are given on the last
  digits of the central prediction when significant. Seven-point scale
  uncertainty is also provided for the cross sections as sub- and
  superscripts in percent. In the last column the polarization fractions  
  are provided.}
\label{tab:xs_fr}
\end{bigcenter}
\end{table}

For the unpolarized cross section, NLO QCD corrections amount to $6.4\%$ compared to the LO result. 
This value is much smaller compared to the $80\%$ correction for the $W^+Z$ process \cite{Denner:2020eck,Le:2022lrp}, 
and $35\%$ for the $ZZ$ case \cite{Denner:2021csi}. 
The smallness of the NLO QCD correction in the $W^+W^-$ process is due to the jet veto, which suppresses the quark-gluon induced 
contribution. 

The jet veto is needed to reduce the top-quark backgrounds, but it increases the theoretical uncertainty \cite{Stewart:2011cf}. The scale uncertainties provided in \tab{tab:xs_fr}, calculated using the seven-point method on the exclusive 
cross section, are known to be significantly smaller than the true uncertainties \cite{Stewart:2011cf}. Moreover PDF uncertainties must be added to get the full theoretical errors. Care must therefore be taken when using the scale uncertainties in \tab{tab:xs_fr} to interpret new physics effects. 

The total NLO EW correction ($\bar{\delta}_\text{EW}$) on the NLO QCD result is $-3.8\%$ for the unpolarized cross section, while it 
ranges from $-2.3\%$ to $-4.5\%$ for the polarized ones. The corrections from the $gg$ and $b\bar{b}$ processes are of similar size, 
with the exception of $+15\%$ correction from the $b\bar{b}$ channel for the LL polarization. 
This large effect comes from the top mass of $173\gev$, setting $M_t=0$ reduces the correction to $+2\%$. 
The contribution from the photon-photon process is very small, being less than $2\%$ for all polarizations.

Taking into account all contributions, the LL polarization fraction is $7.4\%$, while the LT and TL fractions are roughly equal around $12\%$. The TT fraction is dominant, about $69\%$, while the interference effect is negligible, being less than $1\%$. 
Even though being very small for the integrated cross section, this interference is non-vanishing because of the kinematic cuts 
placed on the individual decay leptons. When these cuts are removed, the interference is zero \cite{Denner:2020bcz}.  
 
\subsection{Kinematic distributions}
\label{sect:dist}
\begin{figure}[h!]
  \centering
  \begin{tabular}{cc}
  \includegraphics[width=0.48\textwidth]{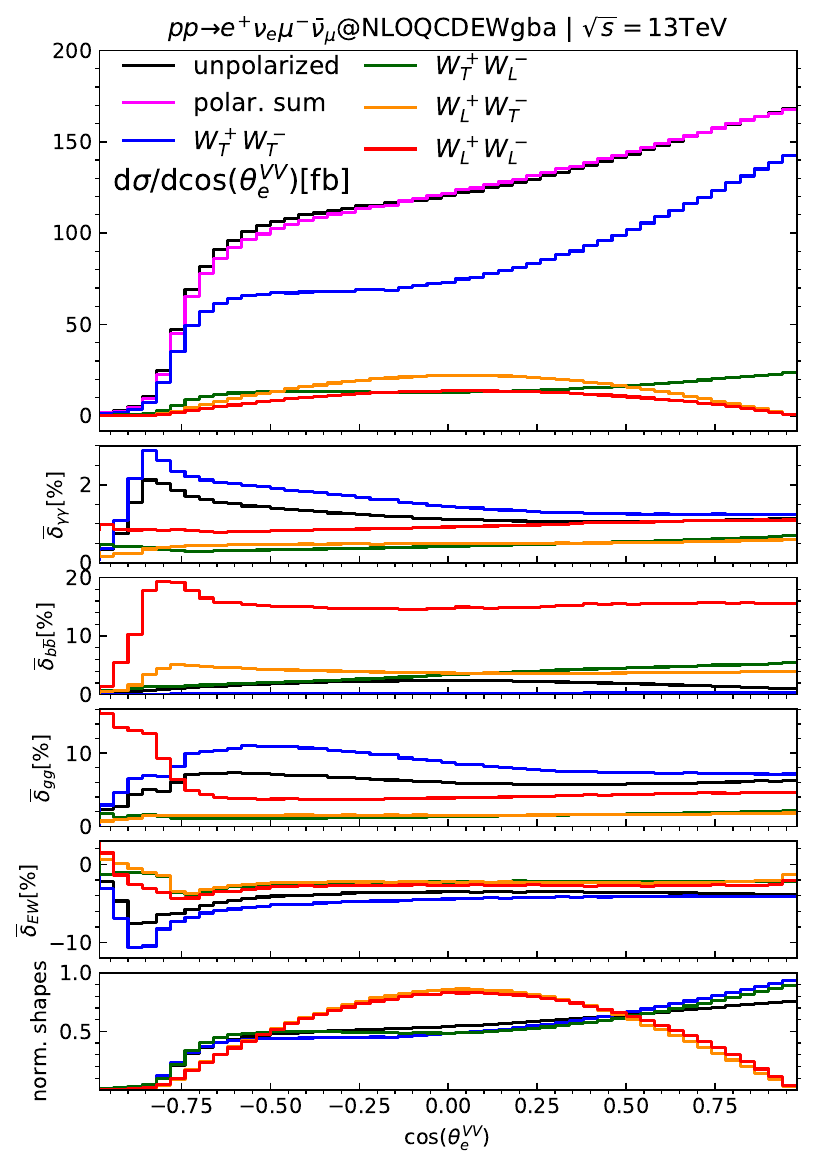} 
  \includegraphics[width=0.48\textwidth]{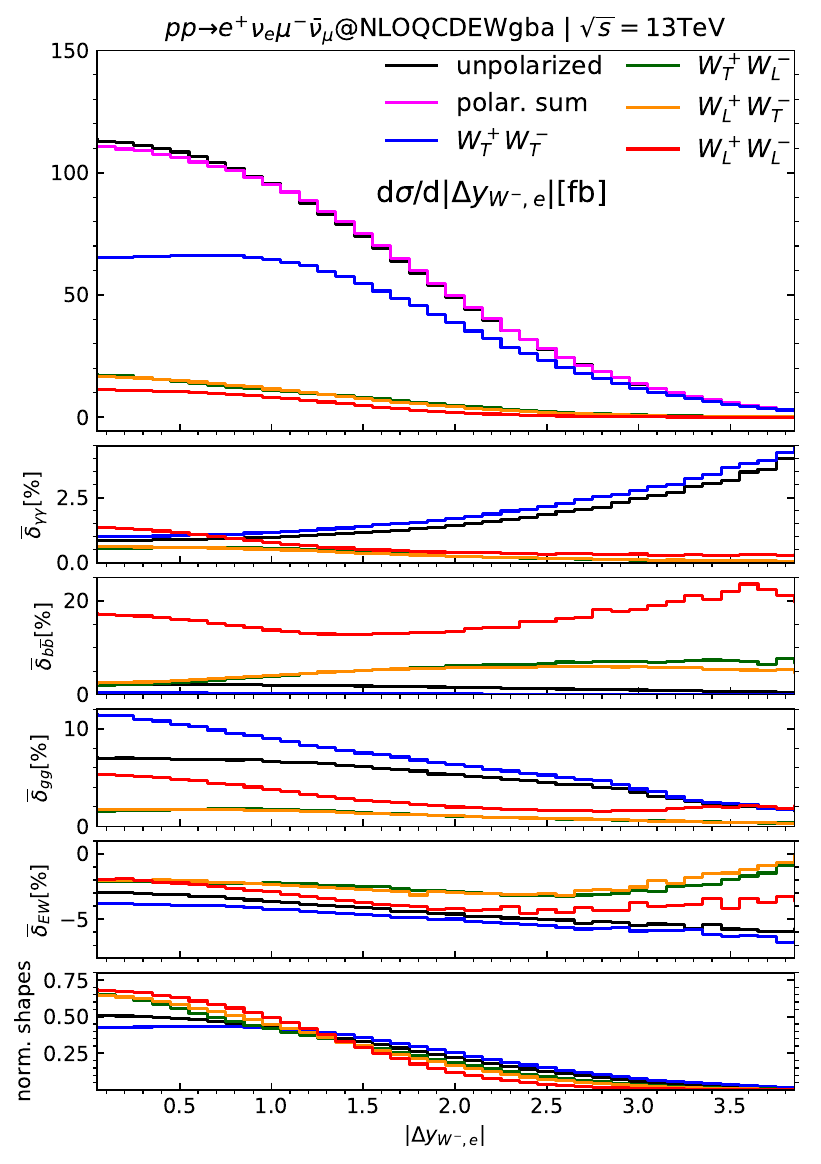}
  \end{tabular}
  \caption{Distributions in $\cos\theta^{VV}_{e}$ (left) and
    $|y_{e^+}-y_{W^-}|$ (right). The big panel
    shows the absolute values of the cross sections including all contributions from 
    the $q\bar{q}$, $gg$, $b\bar{b}$, $\gamma\gamma$ processes. The
    middle panels display the corrections with respect to the NLO QCD $q\bar{q}$ cross sections. 
    The bottom panel shows the
    normalized shapes of the distributions plotted in the top panel.}
  \label{fig:dist_costheta_Deltay}
\end{figure}

We now discuss some differential cross sections which are important for polarization separation. 
The $\cos(\theta_e^\text{VV})$ distribution, where $\theta_e^\text{VV} = \angle(\vec{p}_{e^+},\vec{p}_{W^+})$ with $\vec{p}_{e^+}$ being the positron momentum in the $W^+$ rest frame while $\vec{p}_{W^+}$ the $W^+$ momentum in the VV frame, is presented in \fig{fig:dist_costheta_Deltay} (left). The distribution of rapidity separation between the positron and the $W^-$, $|y_{e^+}-y_{W^-}|$ (calculated in the LAB frame), is shown in \fig{fig:dist_costheta_Deltay} (right). In these plots, the absolute values of the unpolarized and polarized cross sections are shown in the big panel at the top, with all contributions ($q\bar{q}$, $gg$, $b\bar{b}$, $\gamma\gamma$) included. 
In the next four smaller panels below, the corrections $\bar{\delta}_{\gamma\gamma}$, $\bar{\delta}_{b\bar{b}}$, $\bar{\delta}_{gg}$, $\bar{\delta}_\text{EW}$ with respect to the NLO QCD $q\bar{q}$ cross section are plotted. Finally, in the bottom panel, we plot the normalized distributions from the top panel where the integrated cross sections are all normalized to unity, to highlight the shape differences. 

It is found that the EW corrections are mostly negative with the largest magnitude of around 
$10\%$ occurring in the TT polarization at $\cos(\theta_e^\text{VV}) \approx -0.9$. 
For the $b\bar{b}$ correction to the LL polarization, we see that it is consistently large 
over almost the entire phase space. 
It is interesting to observe that photon-photon correction to the TT mode is increasing with large rapidity separation 
$|\Delta y_{W^-,e}|$, while the exact opposite behavior is seen for the gluon-gluon correction. 
The results show that the $gg$, $b\bar{b}$, $\gamma\gamma$ processes contribute differently to different polarization modes 
across different regions of the phase space.     
\begin{figure}[th!]
  \centering
  \begin{tabular}{cc}
  \includegraphics[width=0.48\textwidth]{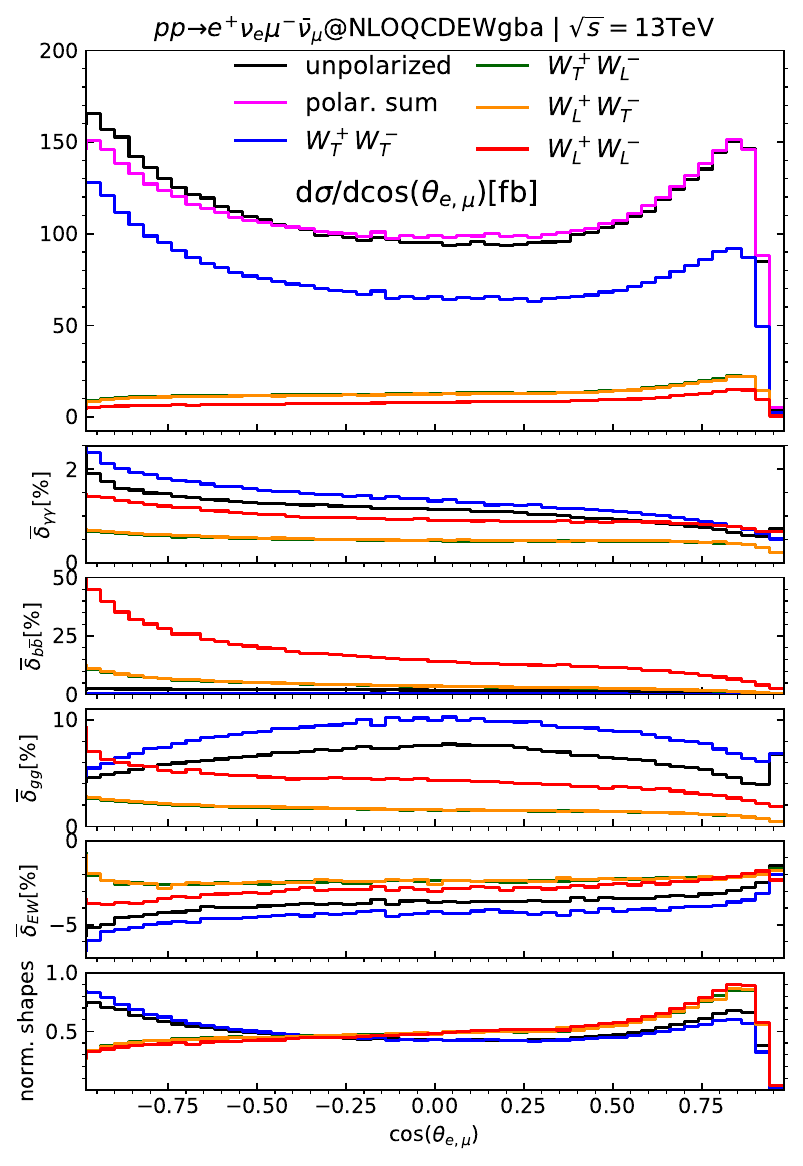} 
  \includegraphics[width=0.48\textwidth]{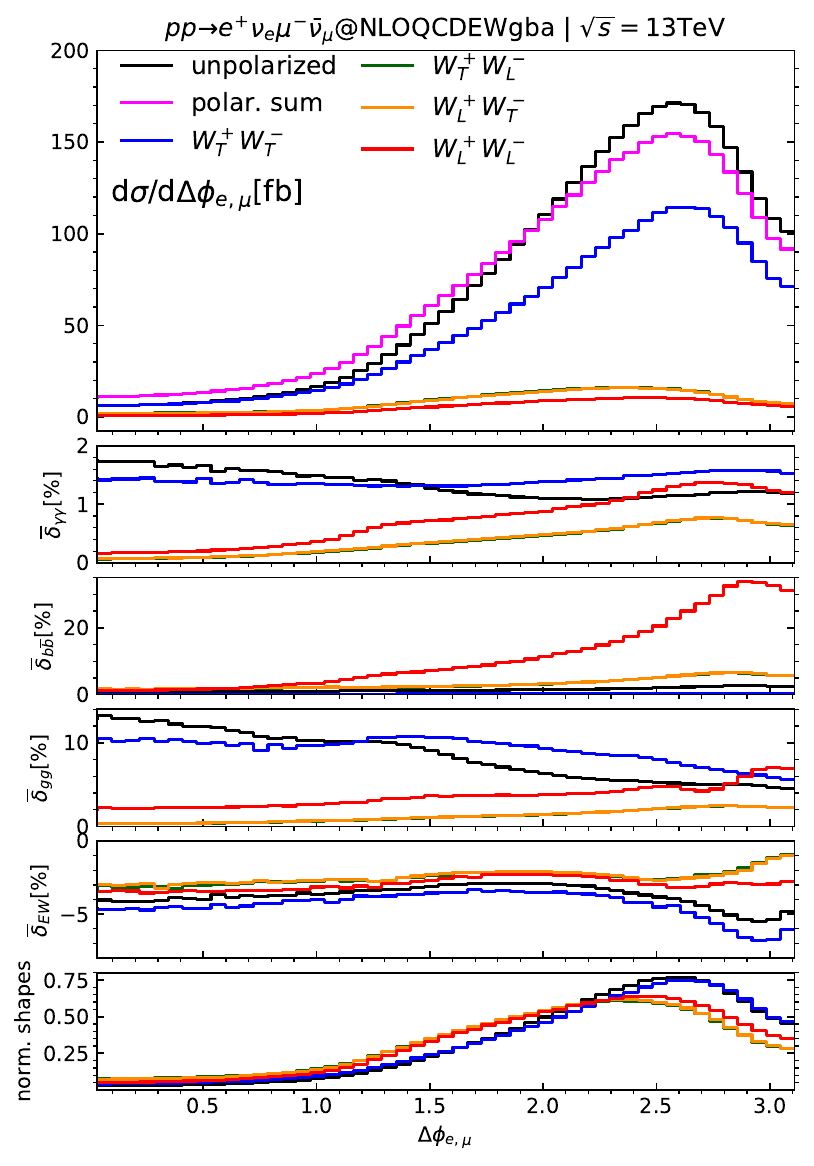}
  \end{tabular}
  \caption{Same as \fig{fig:dist_costheta_Deltay} but for $\cos(\theta_{e,\mu})$ (left) and 
  $\Delta\Phi_{e,\mu}$ (right) distributions.}
  \label{fig:dist_cos_phi_e_mu}
\end{figure}
\begin{figure}[th!]
  \centering
  \begin{tabular}{cc}
  \includegraphics[width=0.48\textwidth]{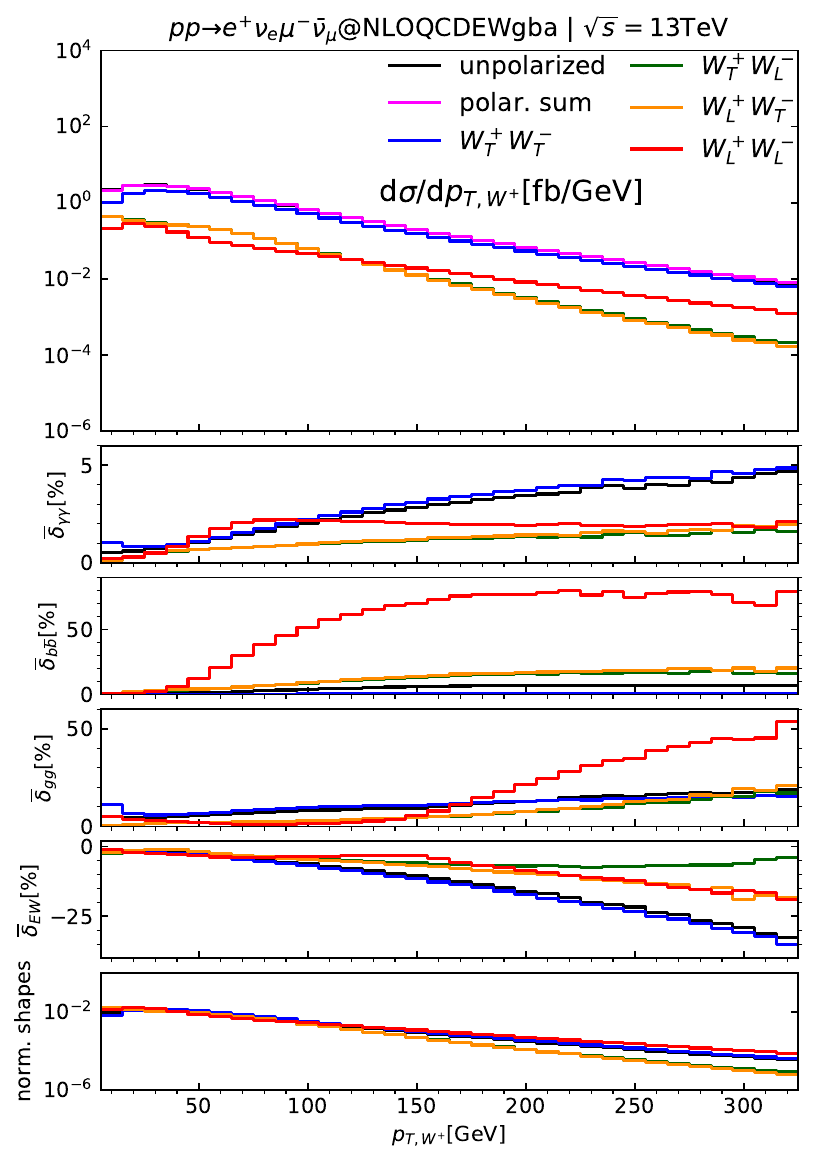} 
  \includegraphics[width=0.48\textwidth]{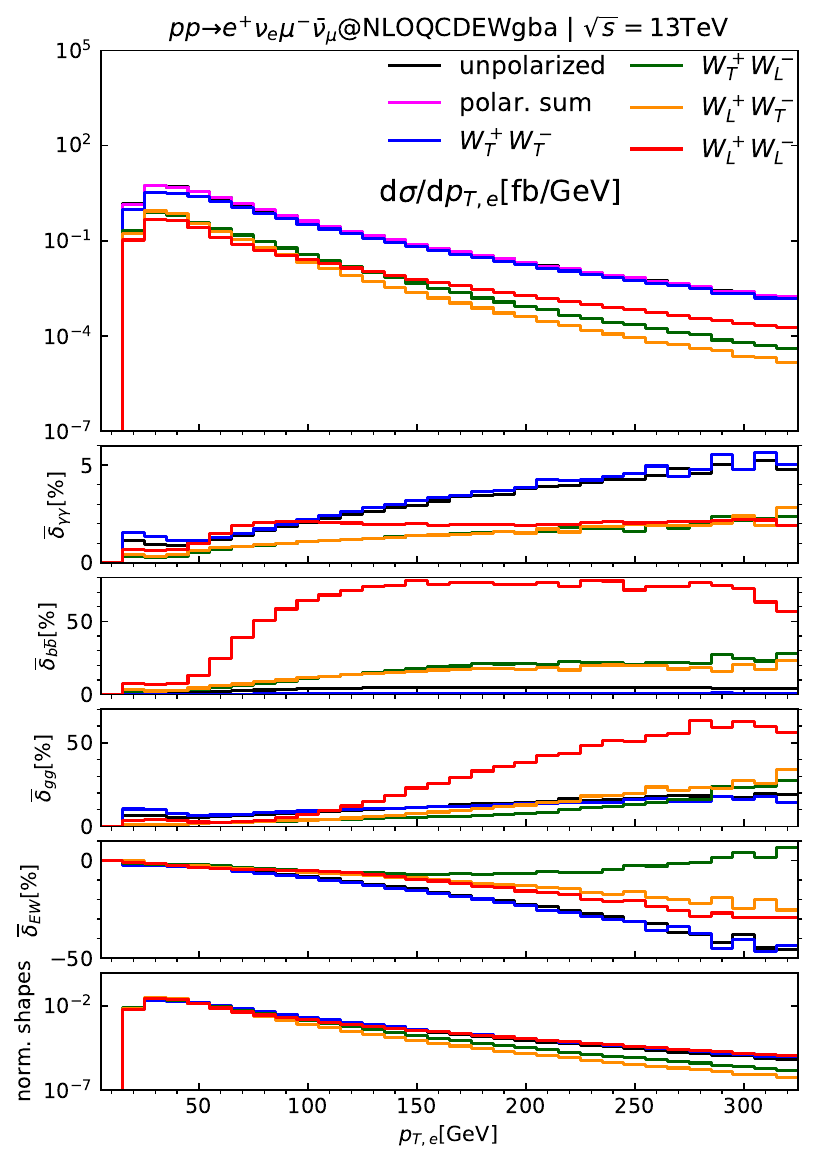}
  \end{tabular}
  \caption{Same as \fig{fig:dist_costheta_Deltay} but for $p_{T,W^+}$ (left) and 
  $p_{T,e^+}$ (right) distributions.}
  \label{fig:dist_pT}
\end{figure}

In similar style, the distributions in $\cos(\theta_{e,\mu})$ with $\theta_{e,\mu}$ being the angle 
between the positron momentum and the muon momentum and in the azimuthal-angle separation $\Delta\Phi_{e,\mu}$ 
are plotted in \fig{fig:dist_cos_phi_e_mu}. The EW corrections are all negative, ranging from $-7\%$ to $-1\%$ 
for all polarization modes. The $b\bar{b}$ correction can reach up to $+45\%$ at $\theta_{e,\mu}=\pi$ 
and up to $+35\%$ when $\Delta\Phi_{e,\mu}\approx \pi$. 

The interference, the difference between the unpolarized (black curve) and the sum of the LL, LT, TL, TT 
cross sections (pink), is sizable and changes sign. 
This effect is more visible in \fig{fig:dist_cos_phi_e_mu} than in the other distributions. 
For the $\cos(\theta_{e,\mu})$ distribution, it is largest, being about $+9\%$ when the 
electron and the muon are maximally separated (i.e. $\theta_{e,\mu}\approx\pi$), and then 
becomes smaller at small separations. 
For the $\Delta\Phi_{e,\mu}$ distribution, the interference effect is much larger, being 
around $-75\%$ when the azimuthal-angle separation is minimal. With increasing separation, 
its magnitude reduces and its sign flips at $\Delta\Phi_{e,\mu}\approx 2.0$~rad, before increasing 
to $+10\%$ at the maximal separation of $\Delta\Phi_{e,\mu}\approx \pi$. 
The interference can be large here because we are looking at the exclusive bins, which is equivalent to 
the situation where kinematic cuts on the individual leptons are applied. 
This large effect should be taken into account when extracting polarization fractions using polarization templates.    

Finally, we present two transverse momentum distributions for the $W^+$ and the positron in \fig{fig:dist_pT}. 
At large $p_T$, the EW corrections are negative and large for $W^+_T W^-_T$, $W^+_L W^-_L$, $W^+_L W^-_T$ modes, 
while being very small for the $W^+_T W^-_L$ (green line). 
The different behavior between the $W^+_L W^-_T$ and $W^+_T W^-_L$ modes 
is because we are looking here at the $p_{T,W^+}$ and $p_{T,{e^+}}$ distributions. 
If we consider the $p_{T,W^-}$ or $p_{T,{\mu^-}}$ distributions (not shown here) 
then the EW corrections are smallest for the $W^+_L W^-_T$ mode. 
It is interesting to notice that the EW corrections are largest for the TT mode. 
In this high $p_T$ region, it is also important to note that contributions 
from the $b\bar{b}$ and $gg$ processes are large for the LL mode.
\section{Conclusions}
\label{sect:conclusion}
In this paper NLO EW corrections to the doubly-polarized cross sections of 
the $W^+W^-$ production process at the LHC have been presented.
The calculation is based on the method provided in our previous work for the $W^\pm Z$ processes \cite{Le:2022ppa}. 
Two new ingredients are needed to complete the dipole subtraction terms, one related to the 
final-state emitter final-state spectator contribution for the photon-radiated process 
and the other related to the photon-photon induced contribution to the quark-photon induced process. 
They do not occur in the $WZ$ process because there is only one charged gauge boson in the final state 
of the on-shell production amplitudes. 
The calculation of these new terms is straightforward following the guidelines of \cite{Le:2022ppa}. 
All details are provided in the appendix.  

Numerical results have been provided for the leptonic $e^+\nu_e \mu^-\bar{\nu}_\mu$ final state using 
a realistic fiducial cut setup inspired from an ATLAS analysis. We found that, in the large $p_T$ region, 
the EW corrections are largest (in absolute value) for the TT polarization mode. 

Besides the main results for the NLO EW corrections, we have also calculated the subleading contributions from the 
$gg$, $b\bar{b}$, $\gamma\gamma$ processes. It was found that the $b\bar{b}$ contribution is important for the 
LL mode, amounting to $15\%$ for the LL integrated cross section, and can be much larger for differential cross sections 
in some regions of the phase space. This effect comes from the heavy top quark. 

\paragraph{Note added:} When we are at the final stage of completing this manuscript, the paper \cite{Denner:2023ehn} performing the same calculation appears. Their numerical results are different from ours because of different kinematic cuts, in particular the absence of a jet veto in their setup. A detailed comparison takes time, hence will be left for future work.     

\appendix

\section{New ingredients in NLO EW corrections}
\label{appen_NLO_cal} 
As mentioned in \sect{sect:pol_NLO}, compared to the $WZ$ calculation 
described in \cite{Le:2022ppa}, there are two new ingredients occurring in 
the $W^+W^-$ process. 
They are the final-state emitter and 
final-state spectator contribution of the OS production part in the dipole-subtraction method and the $\gamma\gamma \to 4l$ 
induced contribution to the quark-photon induced processes occurring at NLO. 

Since these calculations are built on the method of \cite{Le:2022ppa}, 
we need to repeat here the notation to explain them in detail. 
The calculation of NLO EW corrections in the DPA is divided into the 
production and decay parts. The decay part is the same as in the $WZ$ process, hence 
there is no need to repeat it here. The new ingredients are related to the following amplitudes 
of the production part 
\begin{align}
\delta\mathcal{A}_\text{$\gamma$-rad,prod}^{\bar{q}q\to V_1V_2 \to 4l\gamma} &=\fr{1}{Q_1Q_2}
\sum_{\lambda_1,\lambda_2}
\delta\mathcal{A}_\text{$\gamma$-rad,prod}^{\bar{q}q\to
    V_1V_2\gamma}\mathcal{A}_\text{LO}^{V_1\to
    l_1l_2}\mathcal{A}_\text{LO}^{V_2\to l_3l_4},
\label{eq:rad_EW_prod}\\
\delta\mathcal{A}_\text{$\gamma$-ind,prod}^{q\gamma\to V_1V_2q \to 4l q} &=
\fr{1}{Q_1Q_2}\sum_{\lambda_1,\lambda_2}
\delta\mathcal{A}_\text{$\gamma$-ind,prod}^{q\gamma\to
    V_1V_2q}\mathcal{A}_\text{LO}^{V_1\to
    l_1l_2}\mathcal{A}_\text{LO}^{V_2\to l_3l_4},
\label{eq:ind_EW_prod}
\end{align}
where the correction amplitudes
$\delta\mathcal{A}_\text{$\gamma$-rad,prod}^{\bar{q}q\to V_1V_2\gamma}$ ($\gamma$-radiated) and
$\delta\mathcal{A}_\text{$\gamma$-ind,prod}^{q\gamma\to V_1V_2q}$ ($\gamma$-induced) have been
calculated in the OS production calculation in \bib{Baglio:2013toa}
and hence are reused here. The denominators $Q_j$ are computed from the off-shell lepton momenta as in \eq{eq:Qi_def}.

To calculate the corresponding cross sections, we use the dipole-subtraction method \cite{Catani:1996vz,Dittmaier:1999mb} 
where the differential cross section reads
\begin{align}
\left(\fr{d\sigma}{d\xi}\right)_\text{NLO} = &\int d\Phi_n^{(4)} \mathcal{B}(\Phi_n^{(4)}) \delta(\xi - \xi_n) \crn
& + \int d\Phi_n^{(4)} \left[ \mathcal{V}(\Phi_n^{(D)}) + \mathcal{C}(\Phi_n^{(D)}) 
+ \int d\Phi_\text{rad}^{(D)} \mathcal{D}_\text{int}(\Phi_n^{(D)},\Phi_\text{rad}^{(D)}) \right]_{D=4} \delta(\xi - \xi_n) \crn
& + \int d\Phi_{n+1}^{(4)} \left[\mathcal{R}(\Phi_{n+1}^{(4)})\delta(\xi - \xi_{n+1}) - \mathcal{D}_\text{sub}(\tilde{\Phi}_n^{(4)},\Phi_\text{rad}^{(4)})\delta(\xi - \tilde{\xi}_{n}) \right],
\label{Xsection_CS_prod}
\end{align}
where $\mathcal{B}$ and $\mathcal{V}$ are the Born and virtual
contributions.

The amplitudes \eq{eq:rad_EW_prod} and \eq{eq:ind_EW_prod} occur in the $\mathcal{R}$ term in the bottom line of 
\eq{Xsection_CS_prod}. Since they are singular in the IR (soft and collinear) limits, the idea of the subtraction method is to 
introduce a subtraction term $\mathcal{D}_\text{sub}$ which matches the function $\mathcal{R}$ in the singular limits 
so that the new integrand $\mathcal{R}-\mathcal{D}_\text{sub}$ is integrable. 
The corresponding integrated part of $\mathcal{D}_\text{sub}$ is obviously singular, but all these 
divergences cancel in the sum with the virtual corrections ($\mathcal{V}$) and the PDF counter terms ($\mathcal{C}$). 
The $\mathcal{D}_\text{int}$ term in the second line of \eq{Xsection_CS_prod} is a piece of the 
$\mathcal{D}_\text{sub}$ integrated counter part. 
The tilde mark in the bottom line of \eq{Xsection_CS_prod} is to indicate Catani-Seymour mappings, which are 
used to calculate the $n$-particle kinematics from the $(n+1)$-particle kinematics. These mappings are fully provided 
in \cite{Catani:1996vz,Dittmaier:1999mb,Dittmaier:2008md}. The version of \cite{Dittmaier:1999mb} (for the $\gamma$-rad) 
and of \cite{Dittmaier:2008md} (for the $\gamma$-ind) is used in this work.  

The calculation of the $\mathcal{R}$ terms (for both $\text{$\gamma$-rad}$ and $\text{$\gamma$-ind}$) follows the same steps as in \cite{Le:2022ppa} and no complication arises here. 
We first generate the off-shell momenta $[k_{(n+1)}]$ for the $(n+1)$ process. 
From this, a set of OS momenta $[\hat{k}_{(n+1)}]$ is computed using an OS mapping as 
described in \cite{Le:2022ppa}. 

The first new complication arises in the $\mathcal{D}_\text{sub}$ term. For this we need to discuss separately 
the $\text{$\gamma$-rad}$ and $\text{$\gamma$-ind}$ processes.

The $\mathcal{D}_\text{sub}$ term of the $\text{$\gamma$-rad}$ process is a sum of so-called dipole terms. 
Compared to the $WZ$ case, there is a new term called final-state emitter and 
final-state spectator contribution, which happens because both OS $W$ can radiate photon. 
We follow the guidelines of \cite{Dittmaier:1999mb} to calculate this. 
The subtraction function for the OS production $\bar{q} q \to W^+ W^-
\gamma$ reads \cite{Dittmaier:1999mb} (see Section 4.1 there), denoting the OS momenta here as $[p]$ instead of $[\hat{k}]$ 
for simplicity, $p_i=p_{V_i}$, $p_j=p_{V_j}$, $m_i=M_{V_i}$, $m_j=M_{V_j}$ (being the OS $W$ mass here),
\begin{align}
\mathcal{\hat{D}}_\text{sub}^{ij}(p) &\sim \hat{g}_\text{sub}(p_i,p_j,p_\gamma)
\mathcal{\hat{B}}(\tilde{p}_q,\tilde{p}_i,\tilde{p}_j),\label{eq_gOS_ff}\\
\hat{g}_\text{sub} &= \fr{1}{(p_{i} p_\gamma)R_{ij}(\hat{y}_{ij})}\left(\fr{2}{1-\hat{z}_{ij}(1-\hat{y}_{ij})} - 1 - \hat{z}_{ij} - \fr{m_i^2}{p_{i} p_\gamma} \right),\\
\hat{y}_{ij} &= \fr{p_{i}p_\gamma}{p_{i}p_{j} + p_{i}p_\gamma + p_{j}p_\gamma}, \quad 
\hat{z}_{ij} = \fr{p_{i}p_{j}}{p_{i}p_{j} + p_{j} p_\gamma},\\
\tilde{p}_{j}^\mu &= \fr{\sqrt{\lambda(P_{ij}^2,m_i^2,m_j^2)}}{\sqrt{\lambda[(p_{i}+p_\gamma)^2,P_{ij}^2,m_{j}^2]}}\big(p_j^\mu-\fr{P_{ij}p_j}{P_{ij}^2}P_{ij}^\mu\big)+\fr{P_{ij}^2+m_j^2-m_i^2}{2P_{ij}^2}P_{ij}^\mu,\label{p_tilde_1}\\
\tilde{p}_i^\mu &= P_{ij}^\mu - \tilde{p}_j^\mu,\quad P_{ij} = p_i + p_j + p_\gamma,\label{p_tilde_2}
\end{align}
where the subscript $i$ denotes a final-state emitter, $j$ a final-state spectator and the remaining quark momenta $[\tilde{p}_q]$ are the
same as the corresponding quark momenta $[p_q]$, and the following auxiliary functions \cite{Dittmaier:1999mb}
\begin{align}
\lambda(x,y,z) &= x^2 + y^2 + z^2 - 2xy - 2xz - 2yz,\\
R_{ij}(y) &= \frac{\sqrt{[2m_j^2+\bar{P}_{ij}^2(1-y)]^2 - 4P_{ij}^2m_j^2}}{\sqrt{\lambda(P_{ij}^2,m_i^2,m_j^2)}},
\end{align}
with $\bar{P}_{ij}^2 = P_{ij}^2 - m_i^2 - m_j^2$. 
The factor $\mathcal{\hat{B}}(\tilde{p}_q,\tilde{p}_i,\tilde{p}_j)$ in \eq{eq_gOS_ff} is 
the Born amplitude of the reduced process where the photon has been removed. 

Taking into account the leptonic decays using the DPA we have
\begin{align}
\mathcal{\hat{D}}_\text{sub}^{ij}(k)\delta(\xi-\tilde{\xi}_n) &\sim \hat{g}_\text{sub}(\hat{k}_i,\hat{k}_j,\hat{k}_\gamma)
\mathcal{B}(\hat{\tilde{k}}_q,\hat{\tilde{k}}_i,\hat{\tilde{k}}_j)\delta(\xi-\tilde{\xi}_n),\label{eq_gOS_ff_DPA}
\end{align} 
where the on-shell momenta $[\hat{k}_{(n+1)}]$ are obtained from the off-shell momenta $[k_{(n+1)}]$ using an OS mapping as 
in \cite{Le:2022ppa}. We remind that the hat mark is to indicate that the momentum is an OS-mapped quantity.
The singular factor $\hat{g}_\text{sub}$ is calculated using OS momenta as for the 
final-state emitter initial-state spectator terms. Notice that it is Lorentz invariant, hence can be calculated in any reference frame. 
The subtlety comes in the calculation of the reduced Born amplitude. 
For this, we first need to compute the off-shell momenta $[\tilde{k}_{n}]$ from the off-shell momenta $[k_{(n+1)}]$ using CS mapping. 
After this step, an OS mapping is used to calculate $[\hat{\tilde{k}}_{n}]$. 

The off-shell momenta $[\tilde{k}_{n}]$ are obtained as follows. 
We first calculate the off-shell momenta $\tilde{k}_{q}$, $\tilde{k}_{V_1}$, $\tilde{k}_{V_2}$ using 
\eq{p_tilde_1} and \eq{p_tilde_2} with $p_1 = k_e+k_{\nu_e}$, $p_2 = k_\mu+k_{\nu_\mu}$, $m_1^2 = p_1^2$, $m_2^2 = p_2^2$. 
The corresponding tilde momenta for the leptons $\tilde{k}_{e}$, $\tilde{k}_{\nu_e}$ are obtained from $\tilde{k}_{V_1}$ 
using the off-shell mapping described in \cite{Le:2022ppa} (see Eq.~(4.17) there). 
Similarly, $\tilde{k}_{\mu}$, $\tilde{k}_{\nu_\mu}$ are obtained from $\tilde{k}_{V_2}$. 
After this, the same steps as for the final-state emitter and initial-state spectator follow straightforwardly. 

We now come to the $\mathcal{D}_\text{sub}$ term of the $\text{$\gamma$-ind}$ process $q\gamma\to V_1V_2q$ 
with subsequent leptonic decays. There are two singular splittings: $\gamma\to q\bar{q}^{*}$ and $q\to q\gamma^{*}$, 
corresponding to two dipole terms. We follow the method described in \cite{Dittmaier:2008md} to calculate 
these dipole terms. The splitting $\gamma\to q\bar{q}^{*}$, treated in Section 3 of \cite{Dittmaier:2008md}, occurs 
in the $WZ$ process and is similar to the $g\to q\bar{q}^{*}$ splitting in the NLO QCD case. This 
calculation has been described in \cite{Denner:2021csi,Le:2022ppa}. The new thing here is related to the 
$q\to q\gamma^{*}$ splitting, followed by $\gamma\gamma \to W^+W^-$ scattering. 
For this, we follow Section 5 of \cite{Dittmaier:2008md}, using the initial-state spectator 
to calculate the dipole term. Compared to the dipole terms from $W^{*}\to W\gamma$ splitting of the 
$\text{$\gamma$-rad}$ process, the singular factor $g_\text{sub}$ (which is called $h_{\kappa_f,\mu\nu}^{ff,a}$ in Eq. (5.8) of \cite{Dittmaier:2008md}) here is calculated from off-shell momenta 
as the initial-state momenta are not affected by the OS mapping. The flow of the computation reads, denoting the process as
\bea
q(k_q) + \gamma (k_\gamma) \to q (k_{q'}) + l_1 (k_{l_1}) + l_2 (k_{l_2}) + l_3 (k_{l_3}) + l_4 (k_{l_4}).
\label{eq_qa_ind}
\eea 
We first calculate the reduced off-shell momenta $[\tilde{k}_{n}]$ from $[k_{(n+1)}]$ using the following CS mapping \cite{Dittmaier:2008md}:
\begin{align}
\tilde{k}_\gamma &= k_\gamma, \quad \tilde{k}_{\gamma'} = x k_q, \quad \tilde{k}_{l_i}^\mu = \Lambda^\mu_\nu k_{l_i}^\nu,\\
x &= \frac{k_{q}k_\gamma - k_{q'}k_\gamma - k_{q}k_{q'}}{k_{q}k_{\gamma}},\\
\Lambda^\mu_\nu &= g^\mu_\nu - \fr{(P + \tilde{P})^\mu (P + \tilde{P})_\nu}{P^2 + P\tilde{P}} + \fr{2\tilde{P}^\mu P_\nu}{P^2},\\
P &= k_q + k_\gamma - k_{q'}, \quad \tilde{P} = \tilde{k}_{\gamma'} + k_{\gamma}.
\end{align}
The subtraction function $\mathcal{D}_\text{sub}$ is then calculated using Eq. (5.8) of \cite{Dittmaier:2008md} as usual. 
A subtlety one must pay attention to is that the singular function $h_{\kappa_f,\mu\nu}^{ff,a}$ is not Lorentz invariant by itself, hence 
must be calculated in the same reference frame as the reduced Born amplitudes. Needless to say that these reduced amplitudes 
must be calculated using the OS-mapped momenta $[\hat{\tilde{k}}_{n}]$ which are obtained from $[\tilde{k}_{n}]$ using 
a leading-order OS mapping as in \cite{Le:2022ppa}. 

The calculation of the corresponding integrated dipole terms follows straightforwardly as in the 
$WZ$ case, see the guidelines in \cite{Le:2022ppa}. All the needed dipole functions are provided in 
\cite{Dittmaier:1999mb} (for the $\text{$\gamma$-rad}$ case) and in \cite{Dittmaier:2008md} (the $\text{$\gamma$-ind}$).

\acknowledgments
We are grateful to Giovanni Pelliccioli and Ansgar Denner for providing us 
details of their NLO QCD calculation. 
This research is funded by Phenikaa University under grant number PU2023-1-A-18.


\providecommand{\href}[2]{#2}\begingroup\raggedright\endgroup
\end{document}